\documentclass[review=false]{jfp-epi}

\usepackage{mathtools}
\usepackage{listings}
\usepackage{centernot}
\providecommand*{\nlongrightarrow}{\centernot\longrightarrow}

\title{You may implement this later:\texorpdfstring{\\}{ }Cofunctors as partial implementations}

\author{Vincent Wang-Ma\'{s}cianica}
\affiliation{%
  \institution{Human Centered AI Lab, Department of Philosophy, University of Oxford}
  \city{Oxford}
  \country{United Kingdom}
  \authoremail{vincent.wang-mascianica@philosophy.ox.ac.uk}}

\begin{document}
\begin{abstract}
A functor is a familiar model of an implementation, where every operation in a specification is assigned a concrete instantiation at the outset. But some tasks are less eager: we often want to assemble systems while leaving backend choices such as data representations and algorithms for later. We observe that cofunctors admit a direct reading as such \emph{partial implementations}, which are implementations whose extra argument is a state-dependent family of deferred choices. While cofunctors (also called retrofunctors) are not novel, their conceptual and purposive reading in this manner appears to be.
\end{abstract}

\maketitle

\section{Implement now, or later?}\label{sec:eager}

Suppose we have a small language of storage components. At the abstract level we want to write
\[
  \mathsf{Store}
  \xrightarrow{\mathsf{persist}}
  \mathsf{PersistentStore}
  \xrightarrow{\mathsf{replicate}}
  \mathsf{ReplicatedStore}
\]
without immediately deciding whether persistence is provided by SQLite, a remote service, or something else; whether values are serialised as JSON or binary; or which replication protocol is used. This is ordinary software engineering, where closely related ideas are often called \emph{staged configuration}: configuration decisions can be made at different stages, and later choices may depend on earlier ones \citep{czarnecki2005staged}.

Module systems allow implementations to be parameterised and instantiated later \citep{macqueen1984modules}; Standard ML terminology famously and confusingly calls such parameterised modules \emph{functors}. A categorical functor $F : \mathcal S \longrightarrow \mathcal C$ is naturally read as an eager implementation of a specification category $\mathcal S$ in a category of concrete programs $\mathcal C$, where every object $A$ already has an implementation $FA$, and every specification change $f:A\to B$ already has a chosen implementation $Ff:FA\to FB$. Ordinary parameterisation of the ML sort relaxes this to something like $F : \mathcal S \times P \longrightarrow \mathcal C$, where one supplies a global parameter $p:P$ to obtain an implementation.

The interesting case we consider is when a single global parameter type $P$ is not the right shape. For example, the available implementation choices may depend on \emph{where we are in the specification}, where composition may settle old choices and introduce new ones. In the case of the store above, choosing persistence may introduce a serialisation choice, and then choosing replication may add a transaction requirement which the earlier backend either satisfies or exposes as a conflict.

\section{A signature for partial implementations}\label{sec:signature}

Let's first sketch a signature for the problem setting in broad strokes. Suppose $\mathcal S$ is a category of specifications and specification changes. A morphism $f:A\to B$ in $\mathcal S$ is a structured change of specification, such as adding persistence. The operations to be implemented and their laws live inside $A$ and $B$, and $\mathcal S$ need not be a category of host-language types. Let $\mathcal{P}$ be a category of partial implementations and their changes. Elaborating their relationship should get us where we need to go.

On a first approximation, let's say that partial implementations are specifications along with possibly incomplete \emph{choices}, those choices being dependent on the current specification. We can achieve this by an object-mapping $I(A,c)\in\operatorname{Ob}(\mathcal P)$ which selects partial implementations, the objects of $\mathcal P$, given a specification $A \in \operatorname{Ob}(\mathcal S)$ and a choice $c \in \mathsf{Choice}(A)$. What's a choice? Let's associate to every specification object a type of implementation choices $\mathsf{Choice} : \operatorname{Ob}(\mathcal S) \longrightarrow \mathbf{Set}$, where $\mathsf{Choice}(A)$ encodes partial configuration states: some fields may be fixed, some open, and a state may record that accumulated obligations conflict.

Now we have an opening. Taking advantage of the category structure of $\mathcal S$, each specification morphism $f:A\to B$ ought to relate $\mathsf{Choice}(A)$ and $\mathsf{Choice}(B)$, and there are two directions to try. Pulling back, $\mathsf{Choice}(B)\to\mathsf{Choice}(A)$, reads a choice for the richer specification as settling one for the poorer, which is fine for finished implementations (a finished replicated store is in particular a store) but not for obligations: the $\mathsf{protocol}$ field only exists once we replicate, and nothing at $\mathsf{Store}$ has a say in it. Obligations get carried forward and new ones get opened, so we push, defining a \emph{transport} of choices $f_* : \mathsf{Choice}(A)\longrightarrow\mathsf{Choice}(B)$. Evidently, we should like that identities transport choices unchanged, and that transports compose. That is, we ask for $\mathsf{Choice}$ to be a functor $\mathcal S\to\mathbf{Set}$.
\begin{align}
  (1_A)_* &= 1_{\mathsf{Choice}(A)}, \label{eq:transport-id}\\
  (g\circ f)_* &= g_*\circ f_*. \label{eq:transport-compose}
\end{align}

Knowing that $\mathsf{Choice}$ is a functor into $\mathbf{Set}$ gives us a category-theoretic breathsaver: our wordy first approximation for partial implementations ``pairs of specification states and dependent-choices $(A \in \operatorname{Ob}(\mathcal{S}),c \in \mathsf{Choice}(A))$" is just the category of elements $\int_{\mathcal{S}} \mathsf{Choice}$: its objects are the pairs $(A,c)$, and its arrows $(A,c)\to(B,f_*c)$ are exactly the specification changes $f$ out of $A$. Is that exactly the $\mathcal{P}$ we want, or can we say anything else? Let's prod by examples.

Suppose we have two artefacts with identical current behaviour, but different up to unresolved obligation labels $c$ and $c'$. Keeping the distinction seems important, or else we would lose the sort of information we want to carry around for composing the obligations of partial implementations. So let's separate concerns: a later semantics functor can forget the labels and just concern itself with behaviour, but we want the labels to remember distinctions that can tell apart implementations at whatever level we care about (maybe we care about what algorithm we use, maybe we don't), so $(A,c)\mapsto I(A,c)$ had better be an injection into $\operatorname{Ob}(\mathcal P)$. Do we also have surjectivity? An object of $\mathcal P$ outside the image of $I$ would be a partial implementation of nothing, which no specification change could ever act on: dead weight in an environment whose whole job is managing changes. So let's say yes to surjectivity and win ourselves a bijective object map $\operatorname{Ob}(\int_{\mathcal{S}} \mathsf{Choice}) \leftrightarrow \operatorname{Ob}(\mathcal{P})$. May as well go for the gold, is $\mathcal P$ simply $\int_{\mathcal S}\mathsf{Choice}$?  Its arrows are all specification changes, but implementations can also change in ways no specification change induces, for instance swapping one algorithm for another that meets the same specification moves between partial implementations over the same specification. We want $\mathcal P$ to have room for such arrows, so we ask only that the objects agree, and that each transport be implemented by some arrow of $\mathcal P$. Hence the comparison functor, should we have one, from $\int_{\mathcal S}\mathsf{Choice}$ to $\mathcal P$ will be bijective-on-objects and nothing stronger. Now let's clarify the relationships between the various morphisms we have at hand.

Return to the natural foothold, transports, which bridge changing specifications and changing partial implementations. A change in partial implementation we identify as a specification change $f:A\to B$ in $\mathcal S$ with respect to a current implementation state $c \in \mathsf{Choice}(A)$, transporting us via $f_*$ to an implementation state $f_*c \in \mathsf{Choice}(B)$. So we also have a choice-parameterised morphism map from specification-changes to partial-implementation-changes $\mathsf{impl}(f,c): I(A,c)\longrightarrow I(B,f_*c)$. An immediately reasonable ask is that if the specification undergoes no changes at $A$, then neither does the implementation, at any implementation state in $\mathsf{Choice}(A)$.
\begin{align}
  \mathsf{impl}(1_A,c) &= 1_{I(A,c)}. \label{eq:implement-id}
\end{align}

Finally, how does the morphism map $\mathsf{impl}$ behave with respect to composition? This consideration exposes our essential requirement, that \emph{a composite implementation is the same as implementing its first stage, letting that stage update the implementation state, and then implementing the second stage using the updated state.}
\begin{align}
  \mathsf{impl}(g \circ f,c)
    &= \mathsf{impl}(g,f_*c) \circ \mathsf{impl}(f,c). \label{eq:implement-compose}
\end{align}

So we have something reminiscent of a functor relating $\mathcal S$ and $\mathcal P$, with a choice-parameterised object map $I$ and a morphism map $\mathsf{impl}$, and one interesting composition rule capturing the intent of partial implementation. In Idris the signature is nearly literal,\footnote{In a direct simply-typed translation, the result type of \texttt{implement} cannot depend on the values \texttt{f} and \texttt{c}. Idris checks the computed endpoint in the displayed type; an index-erased encoding leaves that check to the program.} and the companion file adds the equational laws \eqref{eq:transport-id}--\eqref{eq:implement-compose}, the object bijection and the category laws.
\begin{lstlisting}[language=Haskell]
Choice : SpecObj -> Type

transport :
  SpecHom a b ->
  Choice a ->
  Choice b

implObj :
  (a : SpecObj) ->
  Choice a ->
  PartialObj

implement :
  (f : SpecHom a b) ->
  (c : Choice a) ->
  PartialHom
    (implObj a c)
    (implObj b (transport f c))
\end{lstlisting}

At the level of parameter shape, this construct behaves like a state-dependent generalisation of ML-functors. If $\mathsf{Choice}(A)=P$ for every $A$ and every $f_*$ is the identity, we recover the familiar constant-parameter case $\mathcal S\times P\to\mathcal C$ (with $P$ discrete), so the dependent version is a natural generalisation of parameterised implementation. We can check our intuitions by verifying that having no choices left recovers a categorical functor, by considering the special case where every $\mathsf{Choice}(A)$ is a singleton. In this case, every transport $f_*$ is forced to be the unique map, and consequently $I$ and $\mathsf{impl}$ indeed reduce to an ordinary functor
\begin{align*}
  F(A)=I(A,*),\quad F(f)=\mathsf{impl}(f,*), \quad F(1_A)=\mathsf{impl}(1_A,*)=1_{I(A,*)}\\
  F(g \circ f) = \mathsf{impl}(g \circ f,*) = \mathsf{impl}(g,*) \circ \mathsf{impl}(f,*) = F(g) \circ F(f)
\end{align*}

A simply-typed translation cannot let $\mathsf{Choice}$ vary with $A$ or let the endpoints of $\mathsf{impl}(f,c)$ mention $c$: without dependent types, the dependence on $c$ becomes an invariant of the program.

\section{That was a cofunctor}

Since $(A,c)\mapsto I(A,c)$ is a bijection onto $\operatorname{Ob}(\mathcal P)$, equations (1)--(4) define a cofunctor $\mathcal S \nlongrightarrow \mathcal P$. This bijection is part of the cofunctor data; functors in general need not be bijective on objects. Clarke's pointwise definition has this object bijection and three axioms \citep[Definition~2.4]{clarke2021diagrammatic}, with targets built into the codomain $I(B,f_*c)$ of $\mathsf{impl}(f,c)$; we have unpacked the identity axiom into~\eqref{eq:transport-id} and~\eqref{eq:implement-id}, and the composition axiom into~\eqref{eq:transport-compose} and~\eqref{eq:implement-compose}.

Cofunctors have a slick categorical presentation due to Clarke, as spans
\[
  \mathcal S
  \xleftarrow{p}
  \Lambda
  \xrightarrow{q}
  \mathcal P,
\]
where $q$ is bijective on objects (identity-on-objects, once $\operatorname{Ob}(\mathcal P)$ is identified with the pairs $(A,c)$) and $p$ is a discrete opfibration \citep[Lemma~2.6]{clarke2021diagrammatic}. This form can be unpacked into more familiar programming intuitions. The discrete opfibration $p$ is the dependent menu of implementation states: its fibre over $A$ is $\mathsf{Choice}(A)$, and its unique lifts are exactly the transports $f_*$. Equivalently, $p:\Lambda\to\mathcal S$ corresponds to the functor $\mathsf{Choice}:\mathcal S\to\mathbf{Set}$, and $\Lambda\cong\int_{\mathcal S}\mathsf{Choice}$ \citep[Definitions~2.13--2.14 and Theorem~2.15]{lobski2026layered2}. By itself, $p$ only records how the menu of choices changes; interpreting a lifted request as an implementation artefact is the task of the other leg $q$. Just as we have deduced, the bijectivity-on-objects of $q$ says that every partial implementation is exactly one apex object, an object of $\mathcal P$ is individuated by its obligation state and nothing else, and $q$ does no identifying of its own.

\subsection*{Peanut gallery}

\textsc{Objection of overcomplication:} \emph{``Hang on, do you really need a cofunctor? Is $\mathcal P$ not just $\int_{\mathcal S}\mathsf{Choice}$ under a new name, so that a partial implementation environment is a copresheaf $\mathsf{Choice}$ together with a semantics functor $\int_{\mathcal S}\mathsf{Choice}\to\mathcal C$?"}

\textsc{Reply:} Section~\ref{sec:signature} gave one difference, the arrows of $\mathcal P$ that no specification change induces. Another difference arises in where the codomain $\mathcal C$ enters. A functor $\int_{\mathcal S}\mathsf{Choice}\to\mathcal C$ implements every state at once; it is the eager functor of Section~\ref{sec:eager} with a larger domain. The cofunctor stops one step short: its objects are obligation states and nothing else, and its arrows are implementation changes that compose and can be carried out before any codomain $\mathcal C$ is declared. Every $\int_{\mathcal S}\mathsf{Choice}\to\mathcal C$ is a cofunctor followed by a functor: let $\mathcal P$ have the objects of $\int_{\mathcal S}\mathsf{Choice}$ and, between their images, the arrows of $\mathcal C$. States with the same behaviour become equal only in that later functor. The cofunctor is an environment for making and managing composite partial implementations, and those choices are supplied from outside, where a semantics is the last of them.

\textsc{Objection of overkill:} \emph{``Smells like bull. To write $\mathsf{Choice}$ at all, don't we have to spell out every implementation state that could ever arise?"}

\textsc{Reply:} Partial implementation is more than configuration happening one step at a time; it is deciding which obligations are today's problems and which are tomorrow's, and this does not require the $\mathsf{Choice}(A)$ type to declare a census of all possible states. A record of $n$ optional fields has $2^n$ obligation states and takes $n$ lines. Each choice fibre carries only the obligations its own specification object has, and new specifications only cost their fibres and transports. What does get fixed up front is the range of values inside a choice fibre. A new backend means a new case in every transport that matches on backends, but that is the ordinary expression problem and we are no worse off than ML functors, which also have to fix up front what their parameters provide.

\textsc{Objection of obvious alternative:} \emph{``Haven't you heard? Backpack \citep[Sections~2.2--2.3]{kilpatrick2014backpack} can already merge or fill hole interfaces with by-name mixin linking.  RTFM."}

\textsc{Reply:} This solves the static assembly problem, but it does not supply a transformation of a \emph{running} store when replication is added, nor a law identifying that migration with the composite obtained by making the store persistent first. These are state-indexed implementation arrows that cofunctors do supply. So, cofunctors can handle the store of Section~\ref{sec:eager} even after it's already deployed and holding data.

\textsc{Objection of ordinary practice:} \emph{``Real work gets done without dependent types, and composing implementation choices was never the hard part. Why should I care?"}

\textsc{Reply:} Dependent types make the invariant visible, but they are not ``what it is about". Dependency injection is a classic pattern, and every container that wires a component graph from a configuration is transporting choices along specification changes and composing the results. We don't have to do the bookkeeping with a rich type system, but because the same mathematics is in play, that means law~\eqref{eq:implement-compose} is instead kept by hand in simply-typed languages.

\section{Procrastinating with cofunctors}

Let's go buy an ice-cold demonstration at the store to chill all that haterade. We are on the hook to show that we can manage executable migration plans with no semantic codomain, while avoiding spelling out every possible implementation state, and rather than static assembly, we'll do it live.

To illustrate, suppose the metadata for a running store begins at
\[
 c_0=\{\mathsf{backend}=\mathsf{SQLite},\;
        \mathsf{serialiser}={?},\;
        \mathsf{cache}=\mathsf{LRU}\}.
\]
Transport through $\mathsf{persist}$ preserves the decisions already made and records a durability migration without filling the missing $?$. Transport through $\mathsf{replicate}$ then records the transaction log and replica seeding, introduces the open field $\mathsf{protocol}={?}$, and checks the backend's transaction capability. Schematically,
\[
 c_0\xmapsto{\mathsf{persist}_*}c_1
    \xmapsto{\mathsf{replicate}_*}c_2,
\]
where $c_2$ is more constrained than $c_0$, but still incomplete. Transport is total because $\mathsf{Choice}(A)$ records obligation states, not just successful final configurations. The choice fibre over $\mathsf{ReplicatedStore}$ is just two constructors, one with two optional fields and a \texttt{Conflict}, and the $\mathsf{protocol}$ field doesn't exist in any earlier fibre. The Idris instance transports SQLite to the next state, but transports a non-transactional flat-file backend to an explicit \texttt{Conflict}:
\begin{lstlisting}[language=Haskell]
replicateChoice : StoreChoice Durable -> StoreChoice Distributed
replicateChoice (DurableChoice SQLite serialiser cache) =
  DistributedChoice SQLite serialiser cache Nothing
replicateChoice (DurableChoice FlatFile serialiser cache) =
  Conflict "replication requires a transactional backend"
replicateChoice (Conflict reason) = Conflict reason
\end{lstlisting}
The category of elements contains the corresponding composable lifts
\[
  (\mathsf{Store},c_0)
  \xrightarrow{\mathsf{persist}}
  (\mathsf{PersistentStore},c_1)
  \xrightarrow{\mathsf{replicate}}
  (\mathsf{ReplicatedStore},c_2).
\]
Applying $q$ gives partially implemented program fragments, and the cofunctor laws enforce:
\[
  \mathsf{impl}(\mathsf{replicate}\circ\mathsf{persist},c_0)
  =
  \mathsf{impl}(\mathsf{replicate},c_1)
  \circ
  \mathsf{impl}(\mathsf{persist},c_0).
\]
In the companion program, $\mathsf{impl}$ builds a typed migration plan in a category $\mathcal P$ of such plans  (no category $\mathcal C$ appears in the file; plans compose before anything runs them),  where a plan is a record of its effects on the deployed store, however many passes produce them. The direct plan for persisting and replicating at once is written on its own as a single pass, and the composition law, the record field \texttt{implementComp}, checks case by case on the choice that it has exactly the effects of the staged plan. Each case holds by \texttt{Refl} because both sides compute to the same record. A one-pass plan that refused a flat-file backend without creating the durable store would be rejected, because the staged plan creates the store first.
\begin{lstlisting}[language=Haskell]
storeImplement :
  (change : StoreChange source target) ->
  (choice : StoreChoice source) ->
  Migration
    (storeImplObj source choice)
    (storeImplObj target (storeTransport change choice))
storeImplement Stay choice = NoMigration
storeImplement Persist choice = Steps (persistEffects choice)
storeImplement Replicate choice = Steps (replicateEffects choice)
storeImplement PersistAndReplicate choice = Steps (onePass choice)

onePass (PlainChoice SQLite Nothing cache) =
  { serialiserOpen := True, durableStore := True, transactionLog := True,
    firstReplica := True, protocolOpen := True } noEffects
onePass (PlainChoice FlatFile Nothing cache) =
  { serialiserOpen := True, durableStore := True,
    blocked := Just "replication requires a transactional backend" }
    noEffects

storeImplementComp Replicate Persist (PlainChoice SQLite Nothing _) = Refl
storeImplementComp Replicate Persist (PlainChoice FlatFile Nothing _) = Refl
\end{lstlisting}
The resulting $c_2$ still has two holes. We may later supply them, for example, by refining
\[
 c_2\rightsquigarrow
 \bar c_2=\{\mathsf{backend}=\mathsf{SQLite},\;
              \mathsf{serialiser}=\mathsf{CBOR},\;
              \mathsf{cache}=\mathsf{LRU},\;
              \mathsf{protocol}=\mathsf{Raft}\}.
\]
Refinement isn't a transport, because there's no specification change behind it: $\rightsquigarrow$ isn't an arrow of $\mathcal S$ at all.  In the companion program it's an arrow \texttt{refine} of $\mathcal P$ from $I(\mathsf{ReplicatedStore},c_2)$ to $I(\mathsf{ReplicatedStore},\bar c_2)$, its target computed from the two commits it carries, which $q$ never reaches because $1_{\mathsf{ReplicatedStore}}$ is the only specification change from $\mathsf{ReplicatedStore}$ to itself and it implements to the identity. The laws say nothing about how such arrows interact with transports. Running the file prints  the whole plan, refinement included, then runs it and completes the store at its target:
\enlargethispage{3\baselineskip}
\begin{lstlisting}
Migration plan:
  - record an open serialiser obligation
  - create the durable store
  - open a transaction log
  - seed the first replica
  - record an open protocol obligation
  - commit serialiser := CBOR
  - commit protocol := Raft
Completed SQLite/CBOR/LRU/Raft store: durable, log open, 1 replica
\end{lstlisting}

\section{Deferred notes}

Procrastination is not a novel idea, and neither are any of the ingredients in this note. The only present contribution is a (so far conspicuously absent) conceptual reading that connects the categorical machinery to a natural usage in functional programming. By way of conclusion, I will try to simultaneously outline the related work and offer reasons for the absence of this connection.

Staged configuration, parameterised modules, and dependent types are also established programming ideas, and functional programming already handles them with different techniques, just not cofunctors. The closest is deferring the interpreter, as in tagless-final encodings \citep{carette2009finally}, free constructions, and effect handlers \citep{plotkin2013handling}, which separate a program from its eventual semantics; but a single interpreter, or handler, gets picked at the end for the whole program, so we are back at the $\mathcal S\times P\to\mathcal C$ shape of Section~\ref{sec:eager} with no per-state obligations for composition to rewrite. The absence of cofunctors is perhaps due to an unfortunate false friend in ``cofunctor'': Kmett's Haskell library documentation notes that contravariant functors are ``sometimes referred to colloquially as \texttt{Cofunctor}'' \citep{kmettContravariant},  which is the pullback we turned down in Section~\ref{sec:signature}.

Aguiar introduced cofunctors for internal categories in his 1997 Cornell dissertation \citep[Section~4.2]{aguiar1997internal}; Di Meglio later proposed the more transparent name ``retrofunctor'' \citep[Chapter~7]{dimeglio2022category}. The pointwise data later identified as a cofunctor already appears as the \texttt{Put} component of Johnson and Rosebrugh's delta lenses for update propagation \citep[Definition~1]{johnson2013delta}, and it is from this body of work that Clarke's span presentation \citep{clarke2021diagrammatic} developed. Ahman and Uustalu's update-update lenses \citep[Section~5]{ahman2017taking} are the nearest miss: they are cofunctors by the authors' own admission, where a view update at $\mathsf{get}(s_0)$ gets simulated by a source update at $s_0$, and if we read the view as the specification and the source as the artefact, that simulation is our lift. But the fibre of $\mathsf{get}$ never gets read as a menu of unresolved choices, nor the lift as an implementation, because the concern there is keeping the next source state consistent with the next view state. Delta lenses and related work tend to consider the interaction of cofunctors paired with functors as generalised \texttt{Put}s and \texttt{Get}s, and it is possible that the concerns of database theory and its distance from functional programming inhibited consideration of cofunctors on their own in this form.

However, cofunctors and their close cousins discrete opfibrations have been put to use on their own in monoidal category theory: in Lobski and Zanasi's layered monoidal theories \citep{lobski2026layered1,lobski2026layered2} and in Wang-Ma\'{s}cianica's cofunctor boxes \citep[Chapter~2]{wangmascianica2023string}, they serve as dependent-choice-parameterised analogues of functors. At the Topos Institute, one strand of work identifies categories with polynomial comonads and their morphisms with cofunctors \citep[Chapter~7]{niu2025polynomial}, and another recasts cofunctors as retromorphisms and extends them to retrotransformations between models of double theories \citep{patterson2023retrotransformations}. So they are certainly theoretically useful devices.

Yet to the best of my knowledge, the easily digestible, eminently practical, and demystifying slogan that ``cofunctors are partial implementations'' is new, so here it is.

\end{document}